\documentclass[a4paper,11pt]{article}
\usepackage{pos}
\usepackage{bm}
\usepackage{braket}
\DeclareMathOperator{\SU}{\mathrm{SU}}
\DeclareMathOperator{\dd}{\mathrm{d}}
\DeclareMathOperator{\Cov}{\mathrm{Cov}}

\newcommand{\omegamin}{\omega_{\mbox{\tiny{min}}}}
\newcommand{\taumax}{\tau_{\mbox{\tiny{max}}}}
\newcommand{\Bnorm}{B_{\mbox{\tiny{norm}}}}

\title{The quenched glueball spectrum from smeared spectral densities}

\author[a]{Marco Panero}
\author*[a,b]{Antonio Smecca}
\author[c]{Nazario Tantalo}
\author[d]{Davide Vadacchino}

\affiliation[a]{Department of Physics, University of Turin \& INFN, Turin\\
Via Pietro Giuria 1, I-20125 Turin, Italy}

\affiliation[b]{Department of Physics, Faculty of Science and Engineering, Swansea University (Singleton Park Campus)\\ Singleton Park, SA2 8PP Swansea, Wales, United Kingdom}

\affiliation[c]{University and INFN of Roma Tor Vergata\\
Via della Ricerca Scientifica 1, I-00133, Rome, Italy}

\affiliation[d]{Centre for Mathematical Sciences, University of Plymouth, Plymouth, PL4 8AA, United Kingdom}

\emailAdd{antonio.smecca@swansea.ac.uk}

\abstract{The standard approach to compute the glueball spectrum on the lattice relies on the evaluation of effective masses from two-point correlation functions of operators with the quantum numbers of the desired state. In this work, we propose an alternative procedure, based on the numerical computation of smeared spectral densities. Even though the extraction of the latter from lattice correlators is a notoriously ill-posed inverse problem, we show that a recently developed numerical method, based on the Backus-Gilbert regularization, provides a robust way to evaluate a smeared version of the spectral densities. Fitting the latter to a combination of Gaussians, we extract the masses of the lightest glueball and of its first excitation in the spectrum of the theory. While the preliminary results presented in this contribution are restricted to simulations at finite lattice spacing and finite volume, and for the purely gluonic sector of QCD, they represent the first step in a systematic investigation of glueballs using spectral-reconstruction methods.}

\FullConference{The 40th International Symposium on Lattice Field Theory (Lattice 2023)\\
July 31st - August 4th, 2023\\
Fermi National Accelerator Laboratory\\}

\begin{document}
\maketitle

\section{Introduction}
\label{sec:introduction}

The existence of glueballs, color-singlet states made only of gluons, with no valence quarks, is a remarkable prediction of quantum chromodynamics (QCD). While this prediction stems directly from the existence of cubic and quartic gluon self-interaction terms in the QCD Lagrangian, and from the confining, intrinsically non-perturbative, nature of QCD (or, more generally, of non-Abelian gauge theories) at low energies, it is remarkable that a rigorous proof of the existence of a mass
gap---the mass of the lightest glueball---is still one of the unsolved Millennium Prize Problems of the Clay Mathematics Institute~\cite{Jaffe:2000ne}. In fact, it may be that this problem will \emph{never} be solved~\cite{Cubitt:2015xsa}.

From the theoretical point of view, the study of the nature and properties of glueballs began in the early days of QCD~\cite{Fritzsch:1975tx}, and for long time it was carried out only by means of phenomenological models, such as, for example, the MIT bag model~\cite{Chodos:1974je}, as was done in refs.~\cite{Jaffe:1975fd, Robson:1978iu, Donoghue:1980hw, Hansson:1982dv, Chanowitz:1982qj, Carlson:1982er, Carlson:1984wq}, or the Isgur-Paton model~\cite{Isgur:1983wj, Isgur:1984bm}, inspired by a confining-string picture~\cite{Mandelstam:1974fq}. Alternative phenomenological models that were used to study the glueball spectrum include those based on the definition of an ``effective gluon mass''~\cite{Cornwall:1981zr, Cornwall:1982zn, Hou:1982dy}. More recently, analytical investigations of the glueball spectrum in QCD-like theories were carried out using the conjectured gauge/gravity correspondence~\cite{Maldacena:1997re, Gubser:1998bc, Witten:1998qj}: examples of such studies include those reported in refs.~\cite{Csaki:1998qr, Hashimoto:1998if, Brower:2000rp, Boschi-Filho:2002wdj, Csaki:2006ji, Gursoy:2007er, Colangelo:2007pt, Juknevich:2009ji, Jarvinen:2011qe, Brunner:2015oqa, Li:2013oda}.

Beside these works based on phenomenological models, numerical investigations of the glueball spectrum, starting from the first principles of QCD (or of its purely gluonic sector, namely $\SU(3)$ Yang-Mills theory without matter fields) can be carried out on the lattice. While such studies have been carried out for more than forty years, and by now there exists an extensive literature on the subject (see ref.~\cite{Vadacchino:2023vnc} for a review), it should be noted that there are still many open questions in this area of research. This motivated us to propose a new way to address the determination of the glueball spectrum, based on the reconstruction of spectral densities with the HLT method proposed in ref.~\cite{Hansen:2019idp}, which has been recently applied in a variety of different problems~\cite{Bulava:2021fre, Boito:2022njs, ExtendedTwistedMassCollaborationETMC:2022sta, Gambino:2022dvu, DelDebbio:2022qgu, Lupo:2022nuj, Bonanno:2023ljc, Frezzotti:2023nun, Evangelista:2023fmt}.

After reviewing the general strategy (and the inherent challenges) of conventional glueball-spectroscopy studies on the lattice in section~\ref{sec:overview}, we describe the reconstruction of smeared spectral densities from lattice correlators in section~\ref{sec:smeared_spectral_densities} and present preliminary results in purely gluonic $\SU(3)$ Yang-Mills theory in section~\ref{sec:preliminary_results}. Finally, section~\ref{sec:conclusions} contains a summary and an outlook on future work in this direction.

\section{Overview of glueball-spectroscopy calculations on the lattice}
\label{sec:overview}

\emph{In principle}, the extraction of glueball masses from the lattice is straightforward: given a zero-momentum, local, gauge-invariant operator $\Phi$ constructed only from spatial gauge links, and with the spin ($J$), parity ($P$) and charge-conjugation ($C$) quantum numbers of the desired physical glueball state, the resolution of the identity in the energy eigenstates allows one to express the two-point connected correlation function as
\begin{align}
\label{correlator}
G(a\tau) = \langle \Phi(a\tau)\Phi(0) \rangle_{\mbox{\tiny{conn}}} = \sum_n 
|A_n|^2 \exp(-a\tau\omega_n),
\end{align}
where $A_n=\langle 0 | \Phi(0)|n\rangle$ represents the overlap between the state created by acting with $\Phi(0)$ on the vacuum and the $n$th energy eigenstate, so that the mass of the lightest state in the given channel can be read off from the exponential decay of $G$ at sufficiently large Euclidean-time separations, $a\tau \to \infty$, i.e., it is given by the smallest $\omega_n$ with $A_n\neq0$.

\emph{In practice}, however, such procedure is non-trivial: to begin with, the ``glueball wave function'' is not known a priori, hence one does not know which operator(s) in each $J^{PC}$ channel have the best overlap with the target physical state. In particular, one may expect the operators with the largest overlap with the physical state to be sufficiently smooth, hence they are often constructed using ``smoothened'' link variables. However, the precise details of the optimal smearing/blocking are not known in advance. In fact, it often happens that the operator that provides the best interpolation of a physical glueball corresponds to some non-trivial linear combination of lattice operators, which should be extracted \emph{a posteriori} from lattice data---namely, one should consider a sufficiently large set of lattice operators with the specified quantum numbers, consider the matrix constructed from all of their correlation functions (including off-diagonal ones), and study the generalized eigenvalue problem to properly disentangle the different states. In addition, also the precise identification of the spin quantum number is non-trivial: at every finite value of the lattice spacing $a$, the regularization of the theory on a hypercubic grid breaks the continuous group of rotations down to the finite subgroup of rotations by integer multiples of $\pi/2$. As a consequence, the eigenstates of the lattice Hamiltonian are classified according to the only five irreducible representations of the octahedral point group, the symmetry group of the cube: $A_1$ and $A_2$ (of dimension $1$), $E$ (of dimension $2$), and $T_1$ and $T_2$ (of dimension $3$), the restriction of continuum spin-$J$ representations to the octahedral point group being called ``subduced representations''. A related issue is the fact that, at finite $a$, the $2J+1$ states corresponding to a continuum spin-$J$ representation get mixed among different representations of the lattice symmetry group, their degeneracy being broken by lattice artifacts. In particular, this means that the ground states of spins $J\ge4$ appear as excited states in some representation of the octahedral point group. Besides, certain channels are also affected by mixing with scattering and torelon states (and with isospin-singlet states in full QCD). Finally, the fact that even the lightest glueball state has a relatively large mass, above $1$~GeV, implies that the two-point correlation functions decay quickly as a function of the Euclidean-time separation, and the computation is affected by a bad signal-to-noise ratio; a way to tackle this problem consists in evaluating these correlators on anisotropic lattices, with a finer spacing in the temporal direction. Nevertheless, (taking also into account that, owing to the periodic boundary conditions for gauge fields in the Euclidean-time direction, the exponentials appearing in eq.~\eqref{correlator} are actually replaced by hyperbolic cosines, limiting the maximum separation to half the lattice size) the number of points from which an effective mass can be reliably extracted is rather limited. For all of these reasons, the extraction of glueball masses from the lattice remains a non-trivial problem.

\section{Glueball spectrum from smeared spectral densities}
\label{sec:smeared_spectral_densities}

As an alternative method to extract the glueball spectrum from the lattice, we propose its study from the reconstruction of (smeared) spectral densities.\footnote{A related idea was recently put forward in ref.~\cite{Pawlowski:2022zhh}.} Note that, in principle, perfect knowledge of the spectral density of a theory would provide information not only about the masses of the physical states, but also about their decay widths. The Euclidean correlator defined in eq.~\eqref{correlator} can be written in the K\"all\'en-Lehmann representation as
\begin{align}
\label{Kallen-Lehmann}
G(a\tau)=\int_{\omegamin}^\infty \dd \omega \rho(\omega) \exp(-a\tau\omega).
\end{align}
Extracting the spectral density $\rho(\omega)$ by numerical inversion of the Laplace transform in eq.~\eqref{Kallen-Lehmann} is, however, an ill-posed problem for lattice calculations, in which $G(a\tau)$ is known for a finite number of Euclidean-time separations and is affected by statistical (and systematic) uncertainties. The matter is further complicated by the fact that, in a lattice of finite linear extent $L$, the actual spectral density receives contributions from a combination of Dirac $\delta$ distributions:
\begin{align}
\label{rhoL}
\rho_L(\omega) = \sum_n \frac{|\langle 0 | \Phi(0)|n\rangle|^2}{2\pi \omega_n} \delta\left(\omega- \omega_n(L)\right).
\end{align}
We address the problem by applying a regularization based on the variant of the Backus-Gilbert method proposed in ref.~\cite{Hansen:2019idp}, which gives access to a \emph{smeared} version $\rho^\sigma_L(\omega)$ of the finite-volume spectral density. In a nutshell, this HLT method generalizes the Backus-Gilbert reconstruction~\cite{Backus:1967nao,Backus:1968trp,Backus:1970uit} by having the smearing function as an input of the algorithm. Introducing $K(\omega,\bm{g})=\sum_{\tau=1}^{\taumax} g_\tau(\sigma)\exp(-a\tau\omega)$, the finite-volume smeared spectral density can be written as
\begin{align}
\label{smeared_rhoL}
\rho^\sigma_L(\omega) = \int_0^\infty \dd \omega ~\rho^\sigma_L(\omega) \Delta_\sigma\left(\omega- \omega_n(L)\right) \simeq a\sum_{\tau=1}^{\taumax} g_\tau(\sigma)G(a\tau).
\end{align}
For a given $\sigma$, the $g_\tau$ coefficients are determined by minimizing the functional
\begin{align}
\label{Wfunc}
W_n[\bm{g}] = \frac{A_n[\bm{g}]}{A_n[\bm{0}]} + \lambda B[\bm{g}],
\end{align}
where $A_n[\bm{g}]$ and $B[\bm{g}]$ are defined as
\begin{align}
\label{A_and_B_definitions}
A_n[\bm{g}] = \int_{\omega_0}^\infty \dd \omega ~w_n(\omega) \left| K(\omega,\bm{g}) -\Delta_\sigma\left(\omega- \omega_n(L)\right) \right|,
\quad
B[\bm{g}] = \Bnorm \sum_{\tau_1,\tau_2=1}^{\taumax} g_{\tau_1} g_{\tau_2} \Cov (\tau_1,\tau_2),
\end{align}
with $\Cov (\tau_1,\tau_2)$ denoting the statistical covariance of the correlator. Note that, in the ideal limit of infinitely precise correlators, the method would yield the coefficients that minimize $A_n[\bm{g}]$; conversely, in the presence of errors, the coefficients that one obtains correspond to an ``optimal balance'' between statistical and systematic. The stability of this procedure to extract the $g_\tau$ coefficients can be studied following the strategy proposed in ref.~\cite{Bulava:2021fre}. Finally, the physical spectral function is obtained as
\begin{align}
\label{double_limit}
\rho(\omega) = \lim_{\sigma \to 0} \left( \lim_{L \to \infty} \rho^\sigma_L(\omega) \right);
\end{align}
note that the two limits do not commute.

\section{Preliminary results}
\label{sec:preliminary_results}

We are currently analyzing two ensembles of purely gluonic $\SU(3)$ configurations obtained with the Wilson gauge action on lattices of hypervolume $(L/a)^4=32^4$ at $\beta=5.8941$ and at $\beta=6.0625$; the statistics is approximately $5\times10^3$ configurations for the coarser lattice, and $1.5\times10^4$ configurations for the finer lattice. We are focusing on the two lowest states in the $J^{PC}=0^{++}$ channel (or, more precisely, in the $A_1$ representation of the octahedral point group), for which we can benchmark our results against those obtained (with the conventional techniques described in section~\ref{sec:overview}) in ref.~\cite{Athenodorou:2020ani}.

\begin{figure}
    \centering
    \includegraphics[width=0.7\columnwidth]{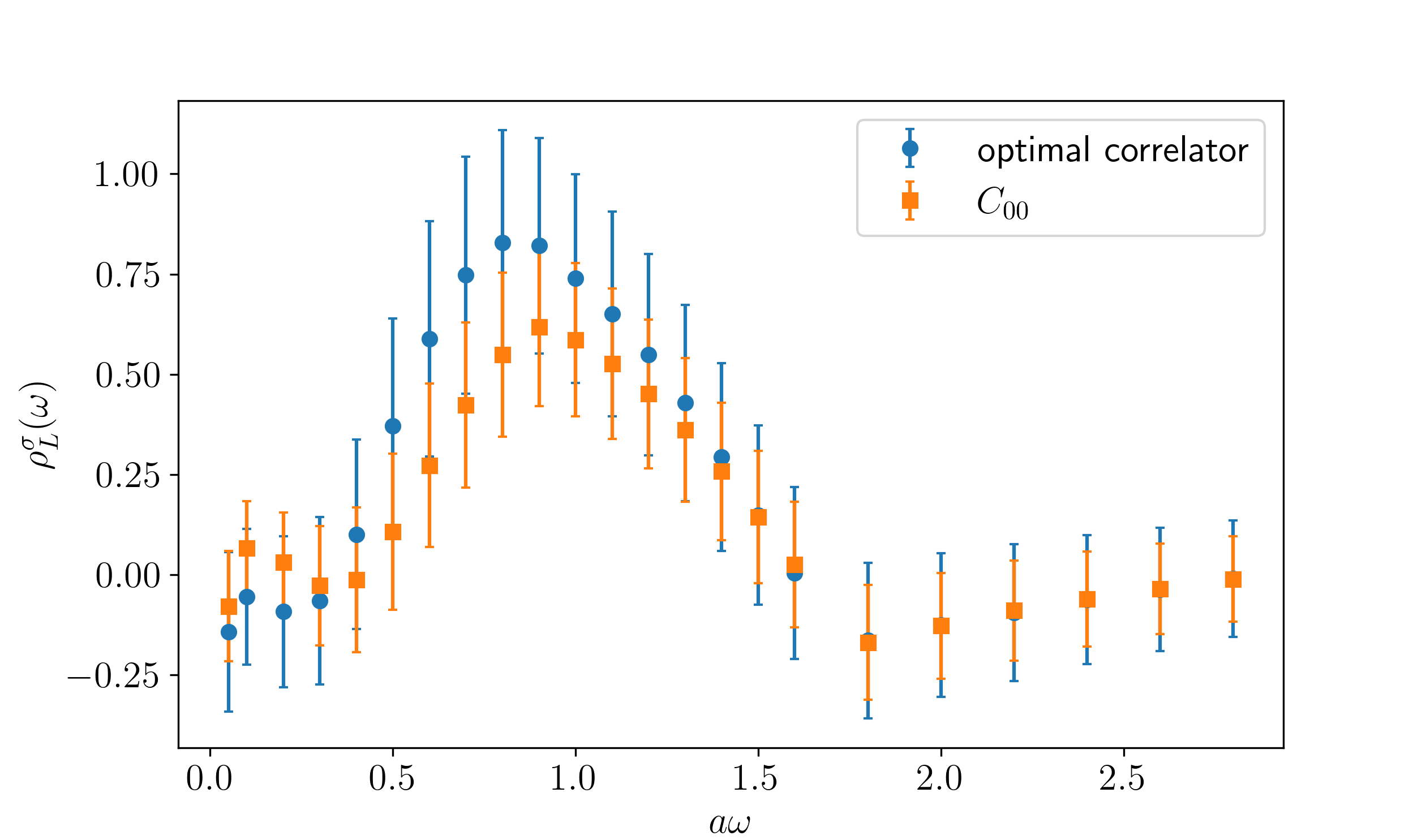}
    \caption{Comparison between the smeared spectral function associated with the correlator derived for the ground state in the $0^{++}$ channel with the variational method (blue symbols) and its contribution from the $C_{00}$ correlator (orange symbols). The data are obtained from simulations at $\beta=5.8941$ and for the value of the smearing parameter in units of the inverse lattice spacing $\sigma a=0.15$.}
    \label{fig:contributions_to_optimal_correlator}
\end{figure}

A first observation that can be made is that our strategy to study the spectral functions allows one to investigate the contributions to the optimal correlators obtained with the variational method: as an example, figure~\ref{fig:contributions_to_optimal_correlator} shows that the spectral function associated with the optimal correlator encoding the propagation of the ground state in the $0^{++}$ channel (at a given lattice spacing and for a fixed smearing parameter $\sigma$) is very close to the one from the correlator $C_{00}$ of the ($0^{++}$ channel projected) combinations of the one-time blocked-smeared plaquette. This is compatible with expectations, since the operators in $C_{00}$ are found to have a sizeable projection on the optimal operators.

\begin{figure}
    \centering
    \includegraphics[width=0.7\columnwidth]{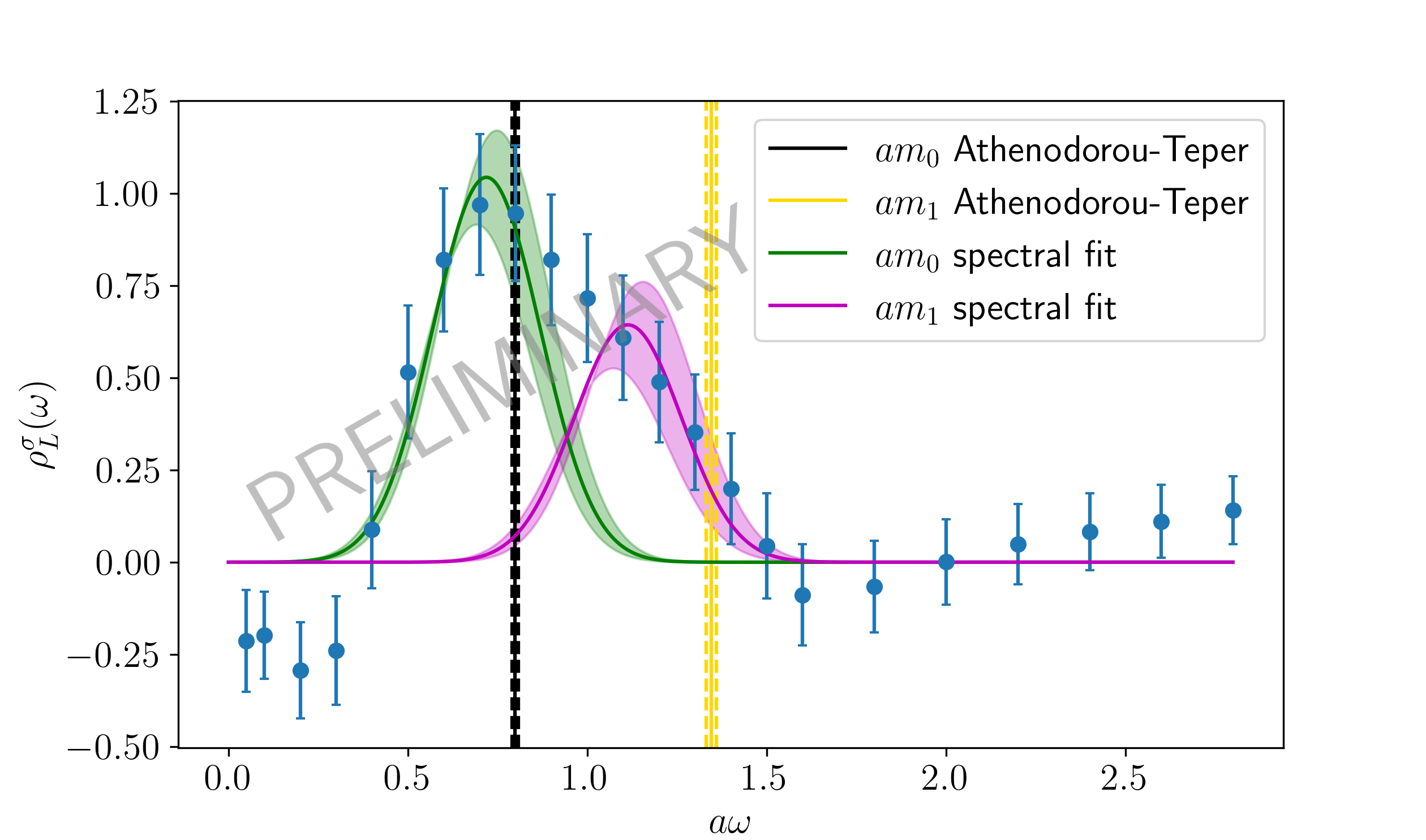}
    \caption{Preliminary results for the fit of the smeared spectral function obtained from our $\beta=5.8941$, $(L/a)^4=32^4$ lattice ensemble with smearing parameter $\sigma a=0.15$ (blue symbols) to a linear combination of two Gaussians (denoted by the green and magenta curves, with the associated uncertainties); the reduced $\chi^2$ of this fit is $2.67$. The black and yellow bars denote the mass of the lightest glueball states in the $0^{++}$ channel estimated in ref.~\cite{Athenodorou:2020ani}.}
    \label{fig:spectral_fit_comparison}
\end{figure}

Next, we remark that, following the approach discussed in refs.~\cite{DelDebbio:2022qgu,Lupo:2022nuj}, it is also possible to directly fit the spectral functions (rather than the correlators), by minimizing the $\chi^2$ defined from the matrix of covariances of the smeared spectral distributions, $\Cov [\rho^\sigma]$. As an example, figure~\ref{fig:spectral_fit_comparison} shows preliminary results for the fit of the smeared spectral distribution obtained from simulations at $\beta=5.8941$ and for $\sigma a=0.15$ to a linear combination of two Gaussians; the reduced $\chi^2$ of this fit is $2.67$. The plot also shows the masses of the two lightest glueballs estimated from a conventional calculation in ref.~\cite{Athenodorou:2020ani}: while this comparison should be taken \emph{cum grano salis} (given that our reconstructed spectral function is at finite $L$ and at finite $\sigma$), the agreement with the results of the variational computation reported in ref.~\cite{Athenodorou:2020ani} appears to be reasonably good.

\begin{figure}[t!]
    \centering
    \includegraphics[width=0.6\columnwidth]{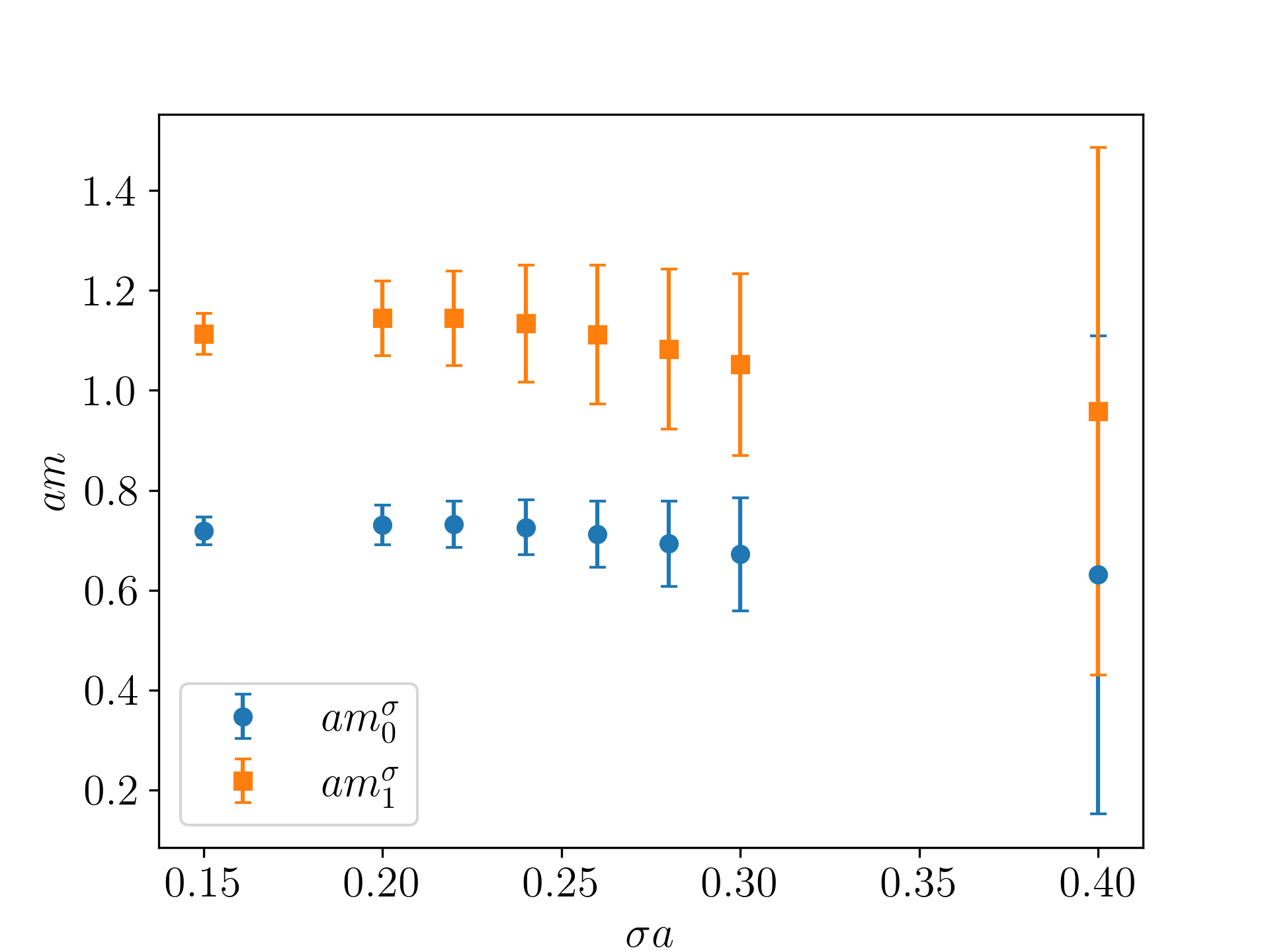}
    \caption{Dependence of our preliminary results for the masses of the two lightest states in the $A_1^{++}$ channel (in units of the inverse lattice spacing) on the smearing parameter $\sigma$ (also in units of $1/a$) from simulations at $\beta=5.8941$, $(L/a)^4=32^4$. The ground state is denoted by blue symbols, while the first excited state is shown by orange symbols.}
    \label{fig:sigma_extrapolation}
\end{figure}

As a matter of fact, one important step in our computation consists in the $\sigma \to 0$ extrapolation of the reconstructed spectral functions. While, as we remarked, this limit should be taken only \emph{after} the extrapolation to the infinite-volume limit, in figure~\ref{fig:sigma_extrapolation} we show a sample of results concerning the dependence of the fitted masses (for both the lightest and the second-lightest states in the $A_1^{++}$ channel) on the smearing parameter---both in units of the lattice spacing. The plot, which refers to preliminary results obtained at $\beta=5.8941$ on a lattice of hypervolume $(L/a)^4=32^4$, reveals that the dependence on $\sigma$ is very smooth, and that the results at small $\sigma$ are affected by small uncertainties. This gives us confidence that the final $\sigma \to 0$ extrapolation of the reconstructed spectral functions will be under control.

\section{Conclusions}
\label{sec:conclusions}

The study of the glueball spectrum remains an interesting and challenging problem in theoretical high-energy physics. In this contribution, we discussed a possible way to address it, by means of the numerical reconstruction of smeared spectral densities, according to the HLT method proposed in ref.~\cite{Hansen:2019idp}. Our preliminary results in purely gluonic $\SU(3)$ Yang-Mills theory are very encouraging, even though, in order to make a sensible comparison with other lattice calculations, a more complete study (including, in particular, the extrapolation to the infinite-volume limit, and the extrapolation to the $\sigma \to 0$ limit) is still required. Thus, the next steps in this ongoing project consist in the improvement of the statistics for our current lattice ensembles, in the generation of new ensembles at larger volume (and at finer lattice spacings, too), and in the investigation of other $J^{PC}$ channels.

\section*{Acknowledgements}
This work was partially supported by the Spoke 1 ``FutureHPC \& BigData'' of the Italian Research Center on High-Performance Computing, Big Data and Quantum Computing (ICSC) funded by MUR (M4C2-19) -- Next Generation EU (NGEU), by the Italian PRIN ``Progetti di Ricerca di Rilevante Interesse Nazionale -- Bando 2022'', prot. 2022TJFCYB, and by the ``Simons Collaboration on Confinement and QCD Strings'' funded by the Simons Foundation. Part of the simulations were run on CINECA computers; we acknowledge support from the SFT Scientific Initiative of the Italian Nuclear Physics Institute (INFN). Part of the simulations were run on the Plymouth HPC cluster. The work of A.~S. is supported by the STFC consolidated grant No.~ST/X000648/1.
The work of D.~V. is supported by STFC in part under the new applicant scheme, and in part under Consolidated Grant No.~ST/X000680/1.


\newpage
\bibliography{paper}
\end{document}